\documentstyle[12pt]{article}

\def\be{\begin{equation}}
\def\ee{\end{equation}}
\def\bdi{\begin{displaymath}}
\def\edi{\end{displaymath}}
\def\br{\begin{eqnarray}}
\def\er{\end{eqnarray}}

\def\u2{\mid u\mid^2}

\def\ra{\rightarrow}

\def\RR{{\rm I\kern-.1567em R}}                              % Doppel R 
 \def\CC{{\rm C\kern-4.7pt                                    % Doppel C 
 \vrule height 7.7pt width 0.4pt depth -0.5pt \phantom {.}}} 
 \def\ZZ{{\sf Z\kern-4.5pt Z}}                                % Doppel Z 

\begin{document}

\begin{titlepage}
\vspace*{-2 cm}
\noindent

\vskip 1cm
\begin{center}
{\Large\bf Integrability from an abelian subgroup of the diffeomorphism
group }

\vglue 1  true cm
C. Adam$^{1a}$,   J. S\'anchez-Guill\'en$^{1b}$,  and 
A. Wereszczy\'nski$^{2c}$
\vspace{1 cm}

$^1${\footnotesize Departamento de F\'\i sica de Part\'\i culas,\\
Facultad de F\'\i sica,
Universidad de Santiago, \, \, and \\
Instituto Galego de Fisica de Altas Enerxias (IGFAE) \\
E-15782 Santiago de Compostela, Spain} \\ ${}$ \\
$^2${\footnotesize Institute of Physics, Jagiellonian University, \\
Reymonta 4,  30-059 Krak\'ow, Poland}

\vspace{1 cm}

\medskip
\end{center}

\normalsize
\vskip 0.2cm

\begin{abstract}
It has been known for some time that for a large class of non-linear field 
theories in Minkowski space with two-dimensional target space the complex
eikonal equation defines integrable
submodels with infinitely many conservation laws. 
These conservation laws are related to the area-preserving diffeomorphisms 
on target space. Here we demonstrate that for all these theories there
exists, in fact, a weaker integrability 
condition which again defines submodels with
infinitely many conservation laws. These conservation laws will be related to
an abelian subgroup of the group of area-preserving diffeomorphisms.
As this weaker integrability condition is much easier to fulfil, it should be
useful in the study of those non-linear field theories.

\end{abstract}

\vfill

$^a${\footnotesize adam@fpaxp1.usc.es} 

$^b${\footnotesize joaquin@fpaxp1.usc.es}

$^c${\footnotesize wereszczynski@th.if.uj.edu.pl}

\end{titlepage}
\section{Introduction }

Recently there has been rising interest in non-linear field 
theories which allow for the existence of knotlike solitons.
The probably best known of these models,
the Faddeev--Niemi model \cite{Fad,FN1}, for example, finds some applications
in condensed matter physics \cite{BFN1,Bab1}. 
Further, some versions of it are discussed 
as possible candidates for a low-energy effective theory of Yang-Mills theory
\cite{FN2,FNW1}.
In addition, there is some intrinsical mathematical interest in 
theories with knot solitons. 
Generally, these models are described by a complex field
$u : \RR^3 \times \RR \to {\cal M}: 
(\vec x ,t) \to u(\vec x ,t)$ where ${\cal M}$ is a two-dimensional target 
space manifold and $u$ plays the role of a complex coordinate on this manifold.

The Faddeev--Niemi model has the two-sphere as target space and is given by
the Lagrangian density
\be \label{FN-L}
{\cal L}_{\rm FN} = {\cal L}_2 - \lambda {\cal L}_4
\ee
where $\lambda $ is a dimensionful coupling constant, 
${\cal L}_2$ is 
\be \label{cp1}
{\cal L}_2 = 4 \frac{\partial_\mu u \, \partial^\mu \bar u}{(1+ u\bar u)^2} ,
\ee
and ${\cal L}_4$ is
\be
 {\cal L}_4 = 4 \frac{(\partial^\mu u \, \partial_\mu \bar u)^2 - (\partial^\mu
u \, \partial_\mu u)(\partial^\nu \bar u \, 
\partial_\nu \bar u)}{(1+u\bar u)^4} .
\ee
Two more models which support solitons and can be constructed
from the two Lagrangian
densities ${\cal L}_2$ and ${\cal L}_4$ separately, are the AFZ 
(=Aratyn, Ferreira and Zimerman) model \cite{AFZ1,AFZ2}
\be
{\cal L}_{\rm AFZ} = -({\cal L}_4)^\frac{3}{4}
\ee
and the Nicole model \cite{Ni1}
\be \label{Ni-La}
{\cal L}_{\rm Ni}= ({\cal L}_2)^\frac{3}{2}.
\ee
Here the noninteger powers for the Lagrangian densities have been chosen
appropriately to avoid Derrick's theorem.
More models together with some explicit
soliton solutions have been constructed, e.g., in 
\cite{Wer1,Wer2}.

Among these models the AFZ model is special, because it has infinitely many 
symmetries and, as a consequence, infinitely many conservation laws
\cite{BF,FR}.
Further, infinitely many soliton solutions can be found by an
explicit integration for a special ansatz (separation of variables in
toroidal coordinates), which realizes the concept of integrability in a
rather explicit way. The other models do not have infinitely many
symmetries, but, nevertheless, ``integrable'' 
subsectors with infinitely many conserved currents
can be defined \cite{AFSG,ASG2}. 
The condition which defines these integrable subsectors
is the complex eikonal equation
\be
u^\mu u_\mu =0
\ee
where $u_\mu \equiv \partial_\mu u$. The infinitely many conserved currents
$J^G_\mu$ (defined in Section 3) for these 
submodels are parametrized by an arbitrary, real function
$G(u ,\bar u)$ and are, in fact, just the Noether currents for the
area-preserving diffeomorphisms on target space \cite{BF,ASG3}. 
[Some more (``generalized'')
integrability conditions, which, however, depend on the Lagrangian,
have been introduced in \cite{Wer3}, \cite{ASG3}.] 

Here we want to demonstrate that there exists, instead of the complex
eikonal equation, a weaker condition which again defines submodels with
infinitely many conservation laws. Further, these integrable
submodels can be defined for
all Lagrangians for which the complex eikonal equation defines integrable
submodels. Explicitly this condition reads
\be \label{int-c}
 \bar u^2 u_\mu^2 - u^2 \bar u_\mu^2 =0.
\ee
The infinitely many conserved currents
$J^G_\mu$ for these submodels are as above, but with the additional 
restriction that now $G=G(u\bar u)$. They are the Noether currents for an
abelian subgroup of the group of  
area-preserving diffeomorphisms on target space.

The meaning of condition (\ref{int-c})
becomes especially transparent when we
re-express $u$ in terms of its modulus and phase like
\be
u=\exp (\Sigma + i \phi ).
\ee
Then the complex eikonal equation is equivalent to the two real equations
\be \label{re-c1}
\Sigma_\mu^2 = \phi_\mu^2
\ee
and
\be \label{re-c2}
\Sigma^\mu \phi_\mu =0  
\ee
whereas the weaker condition (\ref{int-c}) becomes Eq. (\ref{re-c2}) alone
or, for time-independent $u$, 
\be \label{re-c2a}
(\nabla \Sigma) \cdot (\nabla \phi )=0.
\ee

The integrability condition (\ref{int-c}) might be quite useful, for instance, 
in the case of the the Faddeev--Niemi model.
For the Faddeev--Niemi model soliton solutions are only known numerically 
up to now \cite{FN1,GH,BS1,BS2,HiSa}. 
No solutions which solve the complex eikonal equation, as well, 
are known and there are even arguments against the existence of such 
solutions \cite{AFSGW1}.
On the other hand, it is perfectly possible that there exist solutions 
which solve the weaker integrability condition (\ref{int-c}) and that this
condition helps in the search for analytic solutions.

The condition  (\ref{int-c}) is in fact quite weak, i.e., quite easy to
fulfill. For instance, many commonly used separation-of-variable
ans\"atze, like the ansatz $u=\rho (r ,\theta ) e^{im\varphi}$ in
spherical polar coordinates, or the ansatz $u=\rho (\eta ) e^{i(m\varphi 
+ n\xi)}$ in toroidal coordinates (both $\rho $ are real), 
identically obey condition  (\ref{int-c})
due to the orthogonality of the corresponding basis vectors.
On the other hand, for the eikonal equation these ans\"atze 
lead to a differential
equation for the profile function $\rho$ which only allows for very
specific solutions, therefore providing 
a much stronger restriction, see, e.g., 
\cite{Ada1}, \cite{Wer4}.
In short,  condition (\ref{int-c})
applies to a rather large class of field configurations and, therefore, 
we believe that it will be useful for the study
of non-linear field theories with a two-dimensional target space, like
the Faddeev--Niemi or the Nicole model, or the other models mentioned above.

In Section 2 we discuss the algebra of generators of area-preserving 
diffeomorphisms and their abelian subalgebra on a two-dimensional manifold.
Further we define the Noether charges corresponding to these generators.
In Section 3 we show that condition (\ref{int-c}) defines subsectors
with infinitely many conservation laws for a very general class of
Lagrangians (which cover all Lagrangians given above). Further we
demonstrate that the corresponding conserved currents are indeed the Noether 
currents of the abelian area-preserving diffeomorphisms. 

\section{Abelian area-preserving diffeomorphisms}

Here we describe area-preserving diffeomorphisms and an abelian
subgroup contained within them for a two-dimensional manifold 
${\cal M}$ which
later on will be identified with the target space of the non-linear field
theories which we want to study. 
Concretely, we choose real coordinates $(\xi^1 ,\xi^2)$ or
the complex coordinate $u=\xi^1 +i\xi^2$ and allow for the class of 
metrics
\be
ds^2 = g(a ) [(d\xi^1)^2 + (d\xi^2)^2 ] = g(a) du d\bar u
\ee
where 
\be
a=(\xi^1 )^2 + (\xi^2)^2 =u\bar u
\ee
and
\be
du d\bar u \equiv \frac{1}{2} (du \otimes d\bar u + d\bar u \otimes du )
\ee
\be
du \wedge d\bar u \equiv \frac{1}{2} (du \otimes d\bar u - d\bar u 
\otimes du ).
\ee
The corresponding area two-form is
\be \label{ar-fo}
\Omega \equiv g(a) d\xi^1 \wedge d \xi^2 = \frac{g(a)}{2i} d\bar u
\wedge du .
\ee
The choice of conformally flat metrics does not mean a restriction
in two dimensions, because any metric on a two-dimensional manifold
may be chosen conformally flat
by an appropriate choice of coordinates. On the other hand, the 
functional dependence for the metric function $g=g(a)$ is a restriction,
which is however sufficiently general for our purposes. In principle,
one could skip this restriction, which would just complicate the subsequent
discussion without adding substancial new structures (see the remark at the
end of Section 3).  

An area-preserving diffeomorphism 
is a transformation $u\ra v(u,\bar u)$ such that the area form 
(\ref{ar-fo})  remains invariant (see also Refs. \cite{BF}, \cite{FR},
\cite{ASG2}),
\be
\Omega \equiv \frac{1}{2i}g(u\bar u) d\bar u \wedge du =
 \frac{1}{2i}g(v\bar v) d\bar v \wedge dv.
\ee
For infinitesimal transformations $v=u+\epsilon $ it is easy to see that
the condition of invariance of the area form leads to
\be
\epsilon_u +\bar \epsilon_{\bar u} = -\frac{g'}{g}
(\bar u \epsilon +u\bar \epsilon )
\ee
where $\epsilon_u \equiv \partial_u \epsilon$ and $g' \equiv \partial_a g(a)$.
Defining 
\be
\epsilon =g^{-1} \delta \, ,\quad \delta =F_{\bar u}
\ee
the above equation for $\epsilon$ simplifies to
\be \label{Feq}
\partial_u \partial_{\bar u}(F+\bar F)=0.
\ee
The general solution to this equation is
\be \label{Fsol-a}
F +\bar F = \zeta (u) + \bar \zeta (\bar u)
\ee
but for our purposes an imaginary $F$,
\be \label{Fsol} 
F+\bar F=0 ,
\ee
serves as a general solution, because
for any $F$ which solves (\ref{Fsol-a}) there exists a $\tilde F= F-\zeta (u)
$ which is imaginary and leads to the same $\delta =F_{\bar u} =
\tilde F_{\bar u}$, i.e., to the same area-preserving diffeomorphism.

Introducing the real function $G$ via $F=iG$, the area-preserving 
diffeomorphisms are therefore generated by the vector fields
\be
v^G =ig^{-1} (G_{\bar u} \partial_u -G_u \partial_{\bar u})
\ee
which obey the Lie algebra
\be \label{ap-alg}
[v^{G_1},v^{G_2}] = v^{G_3} \, ,\quad G_3 = i g^{-1}
(G_{1,\bar u} G_{2,u} - G_{1,u}G_{2,\bar u}) .
\ee
Now we want to find an abelian subalgebra of this Lie algebra of vector
fields. It is easy to see that the commutator (\ref{ap-alg}) vanishes
if both $G_i, i=1,2$ are of the form 
\be \label{G-ab}
G =G (u\bar u) .
\ee
In addition,
this gives a maximal abelian subalgebra in the sense that if 
$G_1 = G_1 (u\bar u)$ then $ G_3 =0 \, \Leftrightarrow \, 
G_2 = G_2 (u\bar u)$. 
These issues may be seen especially easily by introducing the modulus
and phase of $u$, $u=\sqrt{a} e^{i\phi}$. Then the vector field $v^G$ 
for $G=G(a)$ is
\be
v^G = H(a) \partial_\phi \quad ,\qquad H(a) \equiv g^{-1} G'
\ee
and the above statements follow immediately.
In short, the $G$ of the form $G=G(u\bar u)$ generate
a maximal abelian subgroup of the group of area-preserving diffeomorphisms.

Due to the abelian nature of this subgroup it is trivial to 
integrate the infinitesimal transformations to reach finite ones. The
result is that the transformations
\be
u\ra e^{i\Lambda (u\bar u)} u
\ee
form a subgroup of abelian area-preserving diffeomorphisms, where 
$\Lambda = \Lambda (a)$ is an arbitrary function of its argument. In fact,
these transformations leave invariant the two terms $g(a)$ and
$d\bar u \wedge du$ separately. 

Finally, let us describe how these transformations are implemented for
field theories. For fields $u : \RR^d \times \RR \to {\cal M}: 
(\vec x ,t) \to u(\vec x ,t)$ the generators of area-preserving diffeomorphisms
are given by Noether charges which are constructed with the help of the 
canonical momenta $\pi$, $\bar\pi$ of the fields 
$u$ and $\bar u$.  Concretely, they read
\be \label{Noe-ch}
Q^G =i \int d^d {\bf x} g^{-1} (\bar \pi G_u -\pi G_{\bar u})
\ee
and act on functions of $u,\bar u, \pi ,\bar\pi$ via the Poisson bracket, where
the fundamental Poisson bracket is (with $x^0 =y^0$)
\be
\{ u({\bf x}),\pi ({\bf y}) \} = \{ \bar u({\bf x}),\bar \pi ({\bf y}) \}   
=\delta^d ({\bf x} - {\bf y})
\ee
as usual. The generators $Q^{G_i}$ close under the Poisson bracket,
$\{Q^{G_1} ,Q^{G_2} \} =Q^{G_3}$ where $G_3$ is as in (\ref{ap-alg}).
Specifically, for $G=G(a)$ they generate the abelian area-preserving 
diffeomorphisms, as above.

\section{Integrable subsectors}

In this section we want to show that for a wide class of Lagrangian densities
integrable subsectors can be defined which have infinitely many conserved
Noether currents which may be related to the abelian diffeomorphisms of
the above section. The discussion in this section in some respect resembles
the discussion in Ref. \cite{ASG3}. However, the integrability condition which 
we shall derive here has not been discussed in that reference.
We introduce the class of Lagrangian densities
\be \label{g-lan}
{\cal L} (u ,\bar u ,u_\mu ,\bar u_\mu ) = {\cal F}(a,b,c)   
\ee
where
\be
a=u\bar u \, ,\quad b=u_\mu \bar u^\mu \, ,\quad c= (u_\mu \bar u^\mu )^2
- u_\mu^2 \bar u_\nu^2 
\ee
and ${\cal F}$ is at this moment an arbitrary real function of its arguments.
That is to say, we allow for Lagrangian densities which depend on the fields
and on their first derivatives, are Lorentz invariant, real, and obey the phase
symmetry $u\ra e^{i\lambda} u$ for a constant $\lambda \in \RR$. 
We could relax 
the last condition and allow for real Lagrangian densities which depend on 
$u$ and $\bar u$ independently, but this would just complicate the subsequent
discussion without adding anything substantial. Further, all models we want
to cover fit into the general framework provided by the 
class of Lagrangian densities 
(\ref{g-lan}), therefore we restrict our discussion to this class. 

The canonical four-momentum for this class of models is
\be
\pi_\mu \equiv {\cal L}_{u^\mu} = \bar u^\mu {\cal F}_b + 2 (u^\lambda \bar
u_\lambda \bar u_\mu - \bar u_\lambda^2 u_\mu ){\cal F}_c 
\ee
and the equation of motion reads
\be \label{eom}
\partial^\mu \pi_\mu ={\cal L}_u = \bar u {\cal F}_a
\ee
together with its complex conjugate. 

We introduce the infinitely 
many currents
\be \label{noet-cu}
J^G_\mu = i f(a)( G_u \bar \pi_\mu  - G_{\bar u}  \pi_\mu ) 
\ee
where $f(a)$ is an arbitrary but fixed real function of its argument.
Further, $G$ is an arbitrary real function of $u$ and $\bar u$, and
$G_u \equiv \partial_u G$.  Comparing with the Noether charge (\ref{Noe-ch})
it is tempting to identify $f=g^{-1}$ and $J^G_\mu$ with the Noether
currents of area-preserving diffeomorphisms, and we will see in a moment 
that for a large subclass of Lagrangian densities
this identification can be made, indeed.

In a first step, let us investigate which conditions make the divergence
of the above current vanish, $\partial^\mu J^G_\mu =0$. 
We find after a simple calculation
\br \label{div-jg}
\partial^\mu J^G_\mu & = & if \left( [( M' \bar u G_u + G_{uu} ) u_\mu^2 
 - ( M' u G_{\bar u} + G_{\bar u\bar u}) \bar u_\mu^2 ] {\cal F}_b  \right.
\nonumber \\
&& \left. + \, (uG_u - \bar u G_{\bar u}) [ M' (b{\cal F}_b + 2 c
{\cal F}_c ) + {\cal F}_a ] \right)
\er
where
\be
M \equiv \ln f
\ee
and the prime denotes the derivative with respect to $a$.

The condition that the second term at the r.h.s. of Eq. (\ref{div-jg}) 
vanishes requires that either
\be \label{cond1}
uG_u - \bar u G_{\bar u} =0
\ee
or
\be \label{cond2}
M' (b {\cal F}_b + 2 c {\cal F}_c ) + {\cal F}_a =0.
\ee
Assuming condition (\ref{cond1}) we find the general solution
\be \label{cond1a}
G(u ,\bar u) =  G (u\bar u) \equiv  G(a) 
\ee
which is exactly equal to the condition (\ref{G-ab}) which restricts the
generators of area-preserving diffeomorphisms to 
the abelian subalgebra.

The condition that the first term at the r.h.s. of Eq.  (\ref{div-jg}) 
vanishes requires that either
\be \label{condFb}
{\cal F}_b =0
\ee
or that 
\be \label{cond3}
[( M' \bar u G_u + G_{uu} ) u_\mu^2 
 - ( M' u G_{\bar u} + G_{\bar u\bar u}) \bar u_\mu^2 ] =0 .
\ee
Condition (\ref{condFb}) may, e.g., be satisfied by assuming ${\cal F}_b
\equiv 0 \, \Rightarrow \, {\cal F} = {\cal F}(a,c)$. It follows that theories
with Lagrangians ${\cal L}= {\cal F}(a,c)$ have infinitely many conserved
currents (\ref{noet-cu}), where $G$ is  restricted to (\ref{cond1a}).   Of the
models mentioned in the Introduction, only the
AFZ model falls into this class. However, the AFZ model also obeys
condition (\ref{cond2}), therefore the restriction (\ref{cond1a}) is
unnecessary and the $J^G_\mu$ are conserved for all $G$.

Alternatively we may make the first term at the r.h.s of Eq. (\ref{div-jg}) 
vanish by imposing Eq. (\ref{cond3}). For an unrestricted $G$ this leads to
a condition on the field $u$,
\be \label{eik-eq}
u_\mu^2 =0 ,
\ee
i.e., the complex eikonal equation, which, therefore, defines a submodel for
which there exist infinitely many conserved currents provided that one of 
the two conditions (\ref{cond1}) or (\ref{cond2}) is imposed, in addition. 

However, by
envoking condition (\ref{cond1a}) we may re-express condition (\ref{cond3}) 
like
\be
(M' G' + G'' )F_b [\bar u^2 u_\mu^2 - u^2 \bar u_\mu^2] 
\ee
and, therefore, we find, instead of the complex eikonal equation, the
weaker integrability condition
\be \label{int-ca}
 \bar u^2 u_\mu^2 - u^2 \bar u_\mu^2 =0,
\ee
i.e., Eq. (\ref{int-c}).
Therefore, for {\em all} Lagrangians ${\cal L}= {\cal F}(a,b,c)$ condition
(\ref{int-ca}) defines submodels which have infinitely many conserved
currents (\ref{noet-cu}), where $G$ is  restricted to (\ref{cond1a}), again. 
All models mentioned in the Introduction belong to this class.

Finally, we want to investigate what happens if we impose condition
(\ref{cond2}), either alternatively or in addition to condition 
(\ref{cond1a}) (we want to remark that condition (\ref{cond2}) is fulfilled 
by all models mentioned in the Introduction). 
Equation (\ref{cond2}) 
can be solved easily by the method of characteristics and has the
general solution
\be \label{sol-cha}
{\cal F}(a,b,c)=  {\cal F} (\frac{b}{f},\frac{c}{f^2}) .
\ee
This solution allows to interpret the Lagrangian in terms of the target 
space geometry and to identify the currents (\ref{noet-cu}) with the
Noether currents of the area-preserving diffeomorphisms of Section 2,
as we want to demonstrate briefly.
Indeed, trading the complex $u$ field for two real target space
coordinates $\xi^\alpha$, $u\ra (\xi^1 ,\xi^2 )$, the expressions on which
$ {\cal F}$ may depend can be expressed as follows. The first term is
\be
\frac{b}{f} = \frac{u_\mu \bar u^\mu}{f} = g_{\alpha \beta} (\xi)
\partial^\mu \xi^\alpha \partial_\mu \xi^\beta
\ee
where $\alpha =1,2$ etc, and the target space metric $g_{\alpha \beta}$ is
diagonal and conformally flat for the coordinate choice 
$\xi^1 ={\rm Re}\, u$, $\xi^2 = {\rm Im}
\, u$, i.e., 
\be
g_{\alpha \beta} = g (a)\delta_{\alpha \beta} \equiv 
f^{-1}\delta_{\alpha \beta} .
\ee
For the
second term we get
\be
\frac{c}{f^2}= \tilde \epsilon_{\alpha \beta} \tilde \epsilon_{\gamma \delta}
\partial^\mu \xi^\alpha \partial_\mu \xi^\gamma \partial^\nu \xi^\beta
\partial_\nu \xi^\delta
\ee
where
\be
\tilde \epsilon_{\alpha \beta}= g \,
\epsilon_{\alpha \beta}  \quad , \quad g = f^{-1} =  {\rm det}^\frac{1}{2}
\, (g_{\gamma \delta}) 
\ee
and $\epsilon_{\alpha \beta}$ is the usual antisymmetric symbol in two
dimensions.
We remark 
that the two terms are different in that the first one, $b/f$, depends
on the target space metric, whereas the second one only depends on the
determinant of the target space metric.  For this class of Lagrangians
the currents (\ref{noet-cu}) are the Noether currents of area-preserving
diffeomorphisms on target space, 
and the condition $G=G(a)$ defines these Noether currents 
for the subgroup of abelian area-preserving diffeomorphisms defined in
Section 2, as announced.

{\em Remark:} The abelian subalgebra spanned by generators of the form 
$G=G(u \bar u )$ is by no way the only
abelian subalgebra that exists for the algebra of vector fields $v^G$ of
Eq. (\ref{ap-alg}). In fact, any subset of $G$ of the form $G(u ,\bar u) =
\tilde G_i [h(u,\bar u)]$ where $h$ is an arbitrary but fixed function forms
an abelian subalgebra, i.e. $[v^{\tilde G_1} ,v^{\tilde G_2} ]=0$. This 
follows from the fact that for an area-preserving diffeomorphism the vector
field $v^{\tilde G}$ has to be perpendicular to the (target space) gradient
of $h$, i.e., it has to point into the direction $h={\rm const}$. Indeed,
\be
v^{\tilde G} h = i\tilde G' ( h_{\bar u} \partial_u - h_u \partial_{\bar u}
) h = i\tilde G' (h_{\tilde u} h_u - h_u h_{\tilde u}) =0.
\ee
However, these abelian subalgebras for $h\ne u\bar u$ do not play a special
role in our discussion, 
i.e., they do not produce new integrability conditions. The reason why
$h=u\bar u$ plays a special role lies in the fact that our metric function
(Weyl factor) $g$ depends on it, $g=g(u\bar u)$. Had we chosen a different
functional dependence $g=g[h(u,\bar u)]$ for the metric function, then the
corresponding generators $ \tilde G_i [h(u,\bar u)]$ of an abelian subalgebra
would define a nontrivial new integrability condition. E.g. in the case
$g=g(\xi^1 ) \equiv g(\frac{u+\bar u}{2})$ we find the integrability condition
\be
u_\mu^2 - \bar u_\mu^2 =0 \qquad {\rm or} \qquad (\xi^1)_\mu (\xi^2)^\mu =0.
\ee
A target space with a metric of the form $g=g(\xi^1)$, however, does not
have the topology of the two-sphere (but rather the topology of $\RR^2$
or of a cylinder). Therefore, the corresponding field theory does not have
a nontrivial Hopf index and, consequently, does not give rise to knot
solitons. In this sense it is, therefore, less interesting. \\ \\ \\
{\large\bf Acknowledgement:} \\
This research was partly supported by MCyT(Spain) and FEDER
(FPA2002-01161), Incentivos from Xunta de Galicia and the EC network
"EUCLID". Further, CA acknowledges support from the 
Austrian START award project FWF-Y-137-TEC  
and from the  FWF project P161 05 NO 5 of N.J. Mauser.
AW gratefully acknowledges support from
the Foundation for Polish Science FNP.

\end{document}